\documentstyle[prl,aps,floats]{revtex}
\begin{document}
\tighten
\preprint{PU-RCG/01-XX, gr-qc/0111013 v2}
\input epsf
\draft
\renewcommand{\topfraction}{0.99}
\renewcommand{\bottomfraction}{0.99}
\twocolumn[\hsize\textwidth\columnwidth\hsize\csname
@twocolumnfalse\endcsname

\title{Conservation laws for collisions of branes (or shells) in
general relativity}
\author{David Langlois$^1$, Kei-ichi Maeda$^{2,3}$ and David Wands$^4$\\[-.7em]~}
\address{$^1$ Institut d'Astrophysique de Paris (CNRS), \\
 98 bis Boulevard Arago, 75014 Paris, France\\[-.7em]~}
\address{$^2$Department of Physics, Waseda University, Okubo 3-4-1,
Shinjuku, Tokyo 169-8555, Japan\\[-.7em]~}
\address{$^3$ Advanced Research Institute for Science and Engineering,
Waseda University, Shinjuku, Tokyo 169-8555, Japan
\\[-.7em]~}
\address{$^4$ Institute of Cosmology and Gravitation, University of
  Portsmouth, Portsmouth, PO1 2EG, United Kingdom\\[-.5em]~}
\maketitle
\begin{abstract}
We consider the collision of self-gravitating $n$-branes in a
$(n+2)$-dimensional spacetime.
We show that there is a geometrical constraint which can be expressed
as a simple sum rule for angles
characterizing Lorentz boosts between branes and
the intervening spacetime regions.
This constraint can then be re-interpreted as either energy or
momentum conservation at the collision.
\end{abstract}

\pacs{PACS numbers: 04.25.-g, 04.50.+h \hfill Preprint
  PU-RCG/01-36, gr-qc/0111013 v2\\
to appear in Physical Review Letters}

\vskip2pc]


\def\beq{\begin{equation}}
\def\eeq{\end{equation}}
\def\B{{\cal B}}
\def\C{{\cal C}}
\def\R{{\cal R}}
\def\N{{\cal N}}


\section{Introduction}
Conservation laws during collisions are a cornerstone of Newtonian
mechanics.  They have been generalized to special relativity and are
ubiquitous in the interpretation of collider experiments. In general
relativity, the question of conservation laws during collisions is
subtler because the colliding (self-gravitating) objects affect
spacetime itself.  Some attention has been paid in the literature to
the collision of shells in general
relativity~\cite{kssm81,wu,dt85,redmount85,bi91,nos93,bf96,in99}.
Only particular cases
however have been considered and the calculations are rather involved.
One can distinguish two types of approaches, one based on (not always
well justified) energy-momentum conservation laws, another based on
geometrical constraints, especially in the case of light-like shells.
The interest for collisions in general relativity has been renewed
very recently in the context of brane-world cosmology by the idea that
our current universe could be the product of a collision of 3-branes
in a five-dimensional spacetime~\cite{kost01,kkl01,bucher01}.

In this Letter, we propose a unified treatment, based on a purely
geometric approach, which considerably simplifies the calculations,
and thus immediately applies to {\it any number} of $n$-branes
in a $D=n+2$ dimensional spacetime. The case $n=2$ corresponds
to shells in standard general relativity, whereas the case
$n=3$ applies to the collision of $3$-branes in the brane cosmology
scenarios.
The simplicity of our treatment  follows from
expressing the geometrical constraint as a sum rule for angles
associated with generalised Lorentz boosts between the branes and the
intervening spacetime regions.

\section{Local motion of a brane}

In an $n+2$ dimensional spacetime, $n$-branes divide the spacetime
into distinct regions. We will consider the simplest case where
each such region is empty and can be described by a metric of the
form \beq ds^2=-f(R)dT^2+ {dR^2\over f(R)} +R^2d\Omega_n^2,
\label{metric} \eeq where the `orthogonal' metric $d\Omega_n^2$
does not depend on either $T$ or $R$. The well-known case of a
Schwarzschild-(anti)-de Sitter spacetime corresponds to
$f(R)=k-(\mu/R^{n-1})\mp(R/\ell)^2$.

A brane at the boundary of this region is described by a
two-dimensional trajectory $(T(\tau), R(\tau))$, where $\tau$ is
the proper time.
If we define the two-dimensional velocity vector
$u^a=\left(\dot T, \dot R\right)$,
where the dot denotes the derivative with respect to $\tau$, then
by definition of the proper time, $u^a$ is normalized so that
$g_{ab}u^a u^b= -f \dot T^2+f^{-1} \dot R^2= -1$. One can make
connection with the formulas of special relativity by introducing
a basis of normalized vectors, ${\bf e_T}= f^{-1/2}{\partial\over
\partial T}$ and ${\bf e_R}=\sqrt{f}{\partial \over\partial R}$.

One can then define a Lorentz factor $\gamma=- {\bf e_T}.{\bf u}$ and
a relative velocity $\beta$, given by $\gamma\beta={\bf
  e_R}.{\bf u}$, which yields
\beq
\label{gamma}
\gamma=\sqrt{1+{\dot R^2\over f}} \, , \quad
\gamma\beta={\epsilon\dot R\over \sqrt{f}}.
\eeq
where $\epsilon=+1$ if $R$ decreases from ``left'' to ``right'',  $\epsilon=-1$
otherwise.
Equation~(\ref{gamma}) characterizes the motion of the brane $\B$ with
respect to an observer at rest in the frame $\R$ defined by
(\ref{metric}).  It is easy to check that this implies the standard
special relativistic formula $\gamma=1/\sqrt{1-\beta^2}$.

At any point along the brane trajectory there is a local 
transformation from the bulk coordinates $T$ and $R$ to the proper
time along the brane, $\tau$, and Gaussian normal coordinate, $\chi$:
\beq
 \left(
\begin{array}{c}
d\tau\cr
d\chi\cr
\end{array}
\right)
= \Lambda (-\alpha)
 \left(
\begin{array}{c}
\sqrt{f}dT \cr
\epsilon {dR / \sqrt{f}}\cr
\end{array}
\right),
\label{coordchange}
\eeq
where $\Lambda(\theta)$ is a two-dimensional Lorentz matrix
\beq
\label{Lambda}
\Lambda(\theta)=
\left(
\begin{array}{cc}
\cosh\theta &  \sinh\theta \cr
\sinh\theta & \cosh\theta \cr
\end{array}
\right) \,,
\eeq
and $\alpha$ in Eq. (\ref{coordchange})
is the Lorentz angle associated with the motion of the
brane with respect to the original coordinate systems $\R$, i.e.
\beq
\label{alpha}
\alpha=\sinh^{-1}(\epsilon\dot R/\sqrt{f}).
\eeq

\section{Junction conditions}
Being of codimension $1$, the worldsheet of each  brane we consider here
will separate the spacetime  in two
disconnected regions: the left region, which we call $\R_-$, and the right
region, which we call $\R_+$, with two metrics of the form (\ref{metric})
on the two sides. The coordinate $R$ must be the same on the two sides
because the orthogonal part of the metric must be continuous.

The junction conditions can be written in the form ~\cite{israel}
\beq
[K_{AB}]=-\kappa^2\left(S_{AB}-{S\over n}g_{AB}\right),
\eeq
where the left hand side is the jump of the extrinsic curvature tensor
across
the brane. $S_{AB}$ is the energy-momentum tensor of the brane, $S$ its
trace, and $\kappa^2$ is the coupling between matter and gravity.
For the orthogonal part, the extrinsic curvature components are
$K_{ij}=(\epsilon/R) \sqrt{f+\dot R^2} g_{ij}$, which in the junction
conditions yields the expression
\beq
\label{junction1} \epsilon_+\sqrt{f_++\dot R^2}-\epsilon_-
\sqrt{f_-+\dot R^2}={\kappa^2\over n} \rho R \,,
\eeq
where $\rho$ is the comoving energy density on the brane. This can
be translated into a Friedmann-like equation inside the brane,
which reads
\beq
\dot R^2={\kappa^4\over 4n^2}\rho^2R^2-{1\over
2}\left(f_++f_-\right) +{n^2\over
\kappa^4\rho^2R^2}\left(f_+-f_-\right)^2. \label{friedmann}
\eeq
The other part of the junction conditions is equivalent to the
usual energy conservation law $\dot\rho+n (\dot R/ R)(\rho+P)=0$,
where $P$ is the pressure.

Let us now study the coordinate transformation that relates the coordinates
$(T_+, R_+)$ to the coordinates $(T_-, R_-)$ (note that, for the 
brane,  $R_-=R_+=R$, as imposed by the continuity of the metric along the 
`orthogonal' directions). Because for both
coordinate systems the metric is of the diagonal form (\ref{metric}),
the coordinate transformation is necessarily given by
\beq
 \left(
\begin{array}{c}
\sqrt{f_+}dT_+ \cr
{\epsilon_+ dR_+/ \sqrt{f_+}}\cr
\end{array}
\right)
= \Lambda(\alpha)
 \left(
\begin{array}{c}
\sqrt{f_-}dT_- \cr
{\epsilon_- dR_-/ \sqrt{f_-}}\cr
\end{array}
\right),
\label{lorentz}
\eeq
where $\Lambda$ is a two-dimensional Lorentz matrix as defined in
Eq.~(\ref{Lambda}) .
If one evaluates the coordinate transformation {\em at the brane}, it is
easy to see that the angle  is given by $\alpha=\alpha_+-\alpha_-$,
where $\alpha_+$ and $\alpha_-$ are the Lorentz angles
associated with the motion of the brane with respect to the coordinate
systems $\R_+$ and $\R_-$ respectively, as defined in
Eq.~(\ref{alpha}).
Intuitively, this result is very easy to understand. It simply means that
to go from the coordinates of the region $\R_-$ to the coordinates of the
region $\R_+$, one must do a (pseudo-)Lorentz transformation, which
is the combination of a Lorentz transformation going from $\R_-$ to
a system where the brane is at rest, with a Lorentz transformation from the
brane system to $\R_+$.

\section{System of several branes}
So far, we have considered only one brane and the two regions
surrounding it. To describe the collision of a system of branes in
general, we now introduce a system of $N=N_{in}+N_{out}$ branes,
consisting of $N_{in}$ ingoing branes colliding simultaneously and of
$N_{out}$ outgoing branes, which are produced by the collision.  These
$N$ branes are separated by $N$ different regions of spacetime, which
are assumed to be empty but can be endowed with different cosmological
constants and Schwarzschild masses.

\begin{figure}[t]
\centering
\leavevmode\epsfysize=5cm \epsfbox{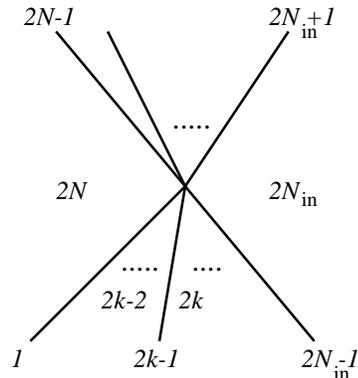}\\
\caption[Nbranes]{\label{Nbranes} Collision of $N_{\rm in}$ ingoing
  branes, yielding $N_{\rm out}=N-N_{\rm in}$ outgoing branes. Odd
  integers denote branes and even integers denote regions in between.}
\end{figure}

To simplify the formalism, we are going to label alternately branes
and regions by integers, starting from the leftmost ingoing brane and
going anticlockwise around the point of collision (see Fig. 1).
 The branes will
thus be denoted by odd integers, $2k-1$ ($1\le k \le N$), and the
regions by even integers, $2k$ ($1\le k \le N$).  Let us introduce, as
before, the angle $\alpha_{2k-1|2k}$ which characterizes
the motion of
the brane $\B_{2k-1}$ with respect to the region $\R_{2k}$,
and which is defined by
\beq
\sinh\alpha_{2k-1|2k}={\epsilon_{2k}\dot R_{2k-1}\over \sqrt{f_{2k}}}.
\label{alpha_k}
\eeq
Of course we can equally describe the motion of the region $\R_{2k}$
with respect to the brane by the Lorentz angle
$\alpha_{2k|2k-1}=-\alpha_{2k-1|2k}$.

We will find it convenient to define a rescaled brane density 
\beq
\tilde \rho_{2k-1}=\pm{\kappa^2\over n}\rho_{2k-1} R, 
\eeq
with the plus sign for ingoing branes ($1\le k\le N_{in}$), the minus 
sign for outgoing branes ($N_{in}+1\le k \le N$).
 An outgoing positive energy density brane
thus has the same sign as an ingoing negative energy density
brane.

The junction condition (\ref{junction1}) then takes the simple
form,
\begin{eqnarray}
\tilde\rho_{2k-1} &=&
\epsilon_{2k}\sqrt{f_{2k}}\cosh\alpha_{2k-1|2k}
\nonumber\\
&& \ - \epsilon_{2k-2}\sqrt{f_{2k-2}} \cosh\alpha_{2k-2|2k-1} \,
\end{eqnarray}
which can be further simplified, using the definition
(\ref{alpha_k}), to give
\begin{eqnarray}
\label{junction2} \tilde\rho_{2k-1} &=&
\epsilon_{2k}\sqrt{f_{2k}}\exp{(\pm\alpha_{2k-1|2k})} \nonumber\\
&& \ - \epsilon_{2k-2}\sqrt{f_{2k-2}}
\exp{(\mp\alpha_{2k-2|2k-1})} \label{rho}.
\end{eqnarray}

\section{Collision and conservation law}
In a small neighbourhood around  the collision event, one can
consider the change of coordinate systems between  two regions in
two ways: going from one region to the next anticlockwise or
clockwise. The requirement
  of having  the same result in the two  cases requires that
the composition of the pseudo-Lorentz transformations must give
identity after a complete tour around the collision event. This
gives the {\it consistency relation} \beq \prod_{k=1}^N
\Lambda(\alpha_{2k-1|2k}-\alpha_{2k-1|2k-2}) = {\bf Id}, \eeq
where we identify the index $i=j+2N$ with $i=j$. This condition
has been obtained recently in a more complicated derivation by
Neronov \cite{neronov01}
using the existence of common null coordinates.
In terms of the Lorentz angles $\alpha$, this consistency relation is
simply the sum rule
\beq
\sum_{i=1}^{2N} \alpha_{i|i+1}=0. \label{collision}
\eeq
This relation provides {\it one constraint}, which can be written
in many ways. What we will show is that this relation can be expressed
in an extremely intuitive form, which can look either like energy
conservation or, equivalently, like momentum conservation.

The main result of this Letter is that, using the junction
conditions~(\ref{junction2}), the sum rule~(\ref{collision}) can be
written as the conservation law
\beq
\sum_{k=1}^N\tilde\rho_{2k-1}e^{\pm
\alpha_{2k-1|j}}=0,
\label{conservation}
\eeq
for {\it any value of the index $j$},
where we have introduced the generalized relative angle
\beq
\label{relativeangle}
\alpha_{j|j'}=\sum_{i=j}^{j'-1}\alpha_{i|i+1},
\eeq
if $j<j'$, and $\alpha_{j'|j}=-\alpha_{j|j'}$.
To prove Eq.~(\ref{conservation}) one can simply use
Eq.~(\ref{junction2}) to substitute for $\tilde\rho_{2k-1}$ and obtain
a sum over exponentials minus another sum which is in fact identical
by Eq.~(\ref{collision}) and hence they cancel each other out.
One must be aware that although one can use Eq.~(\ref{conservation})
to give many different expressions, there is only one underlying
geometrical constraint embodied in (\ref{collision}).

Let us now point out that the conservation law~(\ref{conservation})
can be written as an {\it energy conservation law} seen in the $j$-th
reference frame,
\beq
\sum_{k=1}^N\tilde\rho_{2k-1}\gamma_{j|2k-1}=0,
\eeq
where $\gamma_{j|j'}\equiv \cosh\alpha_{j|j'}$ corresponds to the Lorentz
factor between the brane/region $j$ and the brane/region $j'$ and can be
obtained, if $j$ and $j'$ are not adjacent, by combining all intermediary
Lorentz factors (this is simply using the velocity addition rule of
special relativity), or the relative angle formula~(\ref{relativeangle}).
The index $j$ corresponds to the reference frame with respect to which
the conservation rule is written.

But the  conservation law~(\ref{conservation}) can also be written as
a {\it momentum conservation law} in the $j$-th reference frame,
\beq
\sum _{k=1}^N\tilde\rho_{2k-1}\gamma_{2k-1|j}\beta_{2k-1|j}=0,
\eeq
with $\gamma_{j|j'}\beta_{j|j'}\equiv \sinh\alpha_{j|j'}$.

Note that the relation (\ref{collision}) implies a strong analogy
between the real exponentials (and the hyperbolic cosine and sine),
which we are using here, with the complex exponentials (and the usual
cosine and sine) by effectively imposing a periodicity.

\section{Light-like branes}
Our formalism can also be extended to deal with the case of light-like
branes \cite{bi91}.
One cannot then introduce a local Lorentz transformation from an adjacent
region to the brane frame, as we did above for time-like branes.
But we can still consider  the coordinate
transformation from one of the adjacent regions to the other, which is
still of the form (\ref{lorentz}). Since we now have
$\epsilon_\B dR=\epsilon_+f_+dT_+=\epsilon_-f_- dT_-$  (with
$\epsilon_\B=+1$ for a left-moving brane, $\epsilon_\B=-1$ for a
right-moving brane),
one finds $\epsilon_+=\epsilon_-$ (we have been implicitly assuming here
 that $T$ is a time-like coordinate, i.e. $f>0$,
but it is straightforward  to generalize to the case $f<0$) and
\beq
e^{\epsilon_\B\alpha}= \sqrt{f_-/ f_+}.
\eeq
For example, in the case of two ingoing light-like branes and two
outgoing light-like branes, defining four regions $I$, $II$, $III$ and
$IV$, the substitution of the above result in the sum
rule~(\ref{collision}) immediately yields the DTR
(Dray-t'Hooft-Redmount) formula \cite{dt85,redmount85}
$
f_If_{III}=f_{II}f_{IV}.
$
This can be easily generalized to any combination of time-like branes
with light-like branes, using the general sum rule for angles
(\ref{collision}) and grouping the angles in pairs for the light-like
branes.

\section{Examples}

Let us first consider the case of two ingoing
 branes, $a$ and $b$, colliding to give a single
brane $c$, separated by the regions $I$, $II$ and $III$ (see
 Fig. 2). It is most convenient to express the energy
 conservation law in the frame
of the outgoing brane. One finds
$
\rho_{c}= \rho_a\gamma_{a|c}+\rho_b\gamma_{b|c}.
$
\begin{figure}[t]
\centering
\leavevmode\epsfysize=5cm \epsfbox{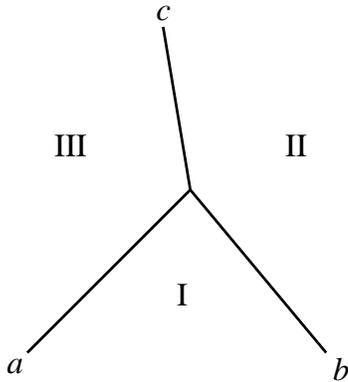}\\
\caption[3branes]{\label{3branes} Collision of two ingoing
  branes, $a$ and $b$, to give single outgoing brane $c$.}
\end{figure}
Note that here, the outcome of the collision is completely determined
by the situation before collision. Indeed, the outgoing brane velocity
and the energy density are interdependent as can be seen in Eq.
(\ref{friedmann}). If one had several outgoing branes then our
conservation law would provide only one relation, which should
completed by other information (based for instance on the microphysics
of the collision) in order to fully determine the outcome.

A slightly more complicated case, but of direct relevance to the
recent ekpyrotic scenario \cite{kost01} or other works on brane
cosmology inspired by the Horava-Witten model~\cite{kkl01}, is when
one of the ingoing branes, $a$ say, is a $Z_2$-symmetric orbifold
fixed point.
We assume that the second brane, $b$, is not $Z_2$-symmetric,
otherwise one dimension of spacetime would disappear
at the collision. Because of the mirror symmetry about $a$, one must
consider two copies, $b$ and $b'$, of the incoming brane (see Fig. 3).
We finally assume that the product of the collision is a single
$Z_2$-symmetric brane, labelled $c$.
Then, $Z_2$-symmetry combined with Eq.~(\ref{collision}) implies that
$\alpha_{a|c}=0$, i.e. there is no redshift between the ingoing and
outgoing $Z_2$-symmetric branes. As a consequence, the energy
conservation law reads simply
$
\rho_c=\rho_a+2\rho_b\gamma_{b|a}.
$
Note that the total momentum is automatically zero in the frame
comoving with $a$ or $c$, however $\dot{R}_c$ is only zero if $\rho_c$
has the critical value given by Eq.~(\ref{friedmann}).
Of course, if one considers the peeling off from an initial $Z_2$-symmetric
brane, $a$, of a brane $b$, then one gets the
conservation law $\rho_a=\rho_c+2\rho_b\gamma_{b|c}$ where the
$Z_2$-symmetric brane after collision is labelled $c$.
Further applications of our results will be discussed in a separate
publication \cite{lmw01b}.

\begin{figure}[t]
\centering
\leavevmode\epsfysize=5cm \epsfbox{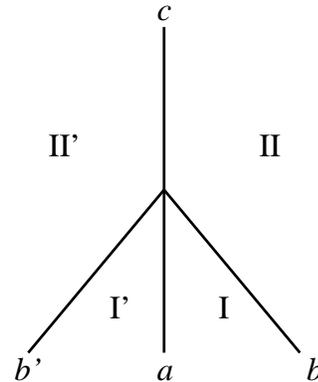}\\
\caption[4branes]{\label{4branes} Z$_2$-symmetric collision. Ingoing
  brane $a$ and outgoing brane $c$ are at orbifold fixed point. Brane
  $b'$ is mirror image of brane $b$.}
\end{figure}

\section{Conclusions}


We have presented here a general treatment of the collision of branes
in a vacuum spacetime.
We have extended to the general case of time-like branes the
geometrical constraint characterizing the collision, which was known
before only in the special case of light-like branes\cite{dt85,redmount85}.
This constraint can be expressed in the simple form of a sum rule for
hyperbolic angles.
The general relativistic junction conditions (\ref{junction2}) allow
us to relate these angles to the energy or momentum of
branes at the collision. 
In this way we obtain extremely simple and intuitive energy and
momentum conservation laws, which are analoguous to the collision of
point particles in two-dimensional special relativity.


%

One can envisage extending our formalism to the case where the
spacetime regions between the branes need be neither empty nor static.
One immediate generalization would be to consider a
Reissner-Nordstrom-(anti-)de Sitter metric in (\ref{metric}). The
formalism would then be unchanged but one would have to supplement the
conservation law (\ref{collision}) with the conservation of the brane
charges. In general, however, the generalization will be complicated
by the need to take into account the junction conditions for the bulk
fields and by the possibility that part of the energy at the collision
might dissipate in excitations of the bulk field.  Nonetheless we
believe that our approach, by its simplicity, is likely to be a useful
starting point for such a generalization.  One application of our
formalism will be to shed some light on the evolution of cosmological
perturbations through a 3-brane collision \cite{lmw01b}.

\begin{acknowledgments}
  We thank C.~Barrabes and J.~A.~Vickers for
  discussions. 
 DL thanks the Portsmouth RCG for its hospitality and KM
 thanks both the IAP and the Portsmouth RCG for their hospitality.
  This work was supported in part by the Yamada foundation. DW is
  supported by the Royal Society.
\end{acknowledgments}


\begin{thebibliography}{99}

\bibitem{kssm81} H. Kodama, M. Sasaki, K. Sato, and K. Maeda,
Prog. Theor. Phys. {\bf 66}, 2052 (1981).

\bibitem{wu}
S.~W.~Hawking, I.~G.~Moss and J.~M.~Stewart,
Phys.\ Rev.\ D {\bf 26}, 2681 (1982);
Z.~C.~Wu,
Phys.\ Rev.\ D {\bf 28}, 1898 (1983).

\bibitem{dt85} T. Dray and G. 't Hooft, Commun. Math. Phys. {\bf 99},
  613 (1985); T. Dray and G. 't Hooft, Class. Quant. Grav. {\bf 3},
  825 (1986). 

\bibitem{redmount85} I.H. Redmount, Prog. Theor. Phys. {\bf 73}, 1401
(1985).

\bibitem{bi91} C. Barrabes, W. Israel, Phys. Rev. D {\bf 43}, 1129 (1991).

\bibitem{nos93} D. Nunez, H.P. de Oliveira, J. Salim, Class. Quantum
Grav. {\bf 10}, 1117 (1993).

\bibitem{bf96} C. Barrabes, V. Frolov, Phys. Rev. D {\bf 53}, 3215 (1996).

\bibitem{in99} D. Ida, K. Nakao, Prog. Theor. Phys. {\bf 101}, 989
(1999).

\bibitem{kost01} 
J.~Khoury, B.~A.~Ovrut, P.~J.~Steinhardt and N.~Turok,
Phys.\ Rev.\ D {\bf 64}, 123522 (2001)
[arXiv:hep-th/0103239].

\bibitem{kkl01}
R.~Kallosh, L.~Kofman and A.~D.~Linde,
Phys.\ Rev.\ D {\bf 64}, 123523 (2001)
[arXiv:hep-th/0104073].

\bibitem{bucher01} 
M. Bucher, ``A braneworld universe from colliding
bubbles'', hep-th/0107148

\bibitem{israel}
W.~Israel, Nuovo Cim. {\bf 44B}, 1 (1966).

\bibitem{neronov01}
A.~Neronov,
JHEP {\bf 0111}, 007 (2001)
[arXiv:hep-th/0109090].

\bibitem{lmw01b} D. Langlois, K. Maeda, D. Wands, in preparation.

\end{thebibliography}
\end{document}